\documentclass[slac_one]{revtex4}
\usepackage{graphicx}
\usepackage{fancyhdr}
\usepackage{subfigure}

\def \Zbb {Zb$\bar{\text{b}}$}
\def \tt {t$\bar{\text{t}}$}
\def \ZZ {ZZ$^{(*)}$}

\begin{document}

\title{{\small{Hadron Collider Physics Symposium (HCP2008),
Galena, Illinois, USA}}\\ 
\vspace{12pt}
Search for the Standard Model Higgs boson decaying to four lepton ($\mu$, e)
final states with the ATLAS experiment at the LHC collider} 

%

\author{Bruno Lenzi for the ATLAS Collaboration}
\email{Bruno.Lenzi@cern.ch}
\affiliation{CEA / Irfu / SPP, Centre de Saclay, F-91191 Gif-sur-Yvette, FRANCE}

\begin{abstract}
The search for the Standard Model Higgs boson in the four lepton
(electron and muon) final state with the ATLAS detector at the LHC is
presented. The analysis strategy and the efficiency for selecting the signal and rejecting the background
are discussed, focusing on the performance of the lepton identification that can be achieved with the first data of LHC expected in 2008.
\end{abstract}

\maketitle

\thispagestyle{fancy}


\section{INTRODUCTION} 

The Higgs boson is the missing piece of the Standard Model (SM) of electroweak and strong interactions. Its mass is a free parameter of the theory, but direct searches at LEP have set a lower bound at 114.4~GeV \cite{Higgs_LEP} and precision electroweak data from LEP and the Tevatron suggest light Higgs scenarios, with $m_H < 199$~GeV at 95\% CL.

The Large Hadron Collider (LHC), a proton-proton accelerator installed in a 27~km tunnel close to Geneva, and two general purpose experiments - ATLAS and CMS - were designed to search for the Higgs boson and evidences of new physics beyond the Stardard Model. The first LHC data is expected within the year 2008 at a center of mass energy of 10~TeV. After the initial machine commissioning, the accelerator should operate at its nominal energy of 14~TeV from 2009 on.

This proceeding describes the discovery potential of the Higgs boson decaying to four leptons - electrons and muons - in ATLAS. This has been recently evaluated with dedicated simulated samples including the signal and corresponding backgrounds produced using the full detector description and reconstruction software.

\section{THE HIGGS BOSON PRODUCTION AND DECAY MODES}

\begin{figure}[!h]
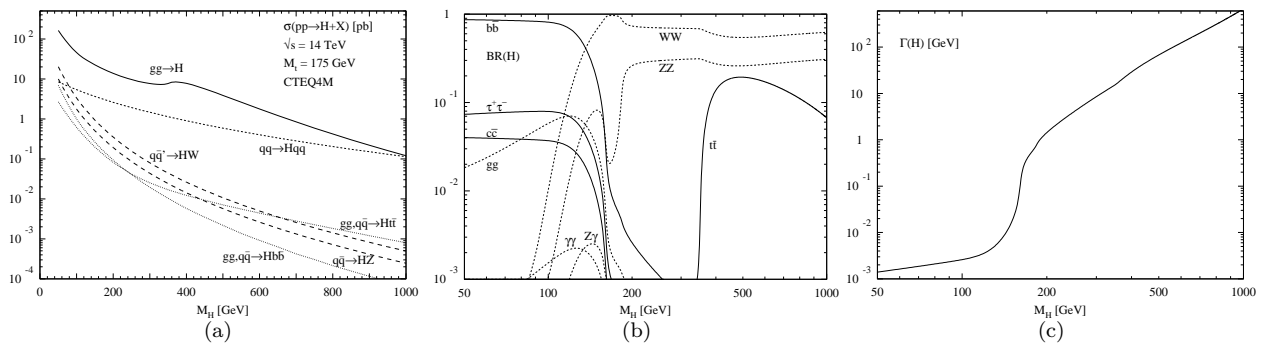

\subfigure[]{\includegraphics[width=0.23\textwidth, angle=-90]{higgs_production.epsi}
 \label{fig:Higgs_production}}
\subfigure[]{\includegraphics[width=0.23\textwidth, angle=-90]{higgs_bratios.epsi}%
\label{fig:Higgs_bratios}}
\subfigure[]{\includegraphics[width=0.23\textwidth, angle=-90]{higgs_width.epsi}%
\label{fig:Higgs_width}}
\caption{(a) Cross-sections of the various Higgs production modes in the LHC \cite{Zerwas}, (b) branching fractions and (c) width for the decay of the Higgs boson as a function of its mass  \cite{Djouadi}. }
\end{figure}

The main production modes for the Higgs boson at the LHC are showed in Fig.~\ref{fig:Higgs_production}. Gluon-fusion is the dominant process, accounting for approximately 80\% depending on the mass, followed by Weak Boson Fusion (WBF). The Higgs decays preferentially to the heaviest particles kinematically accesible - shown in Fig.~\ref{fig:Higgs_bratios}, which means that W and Z bosons start to dominate for masses above 140~GeV. This also implies that the width of the particle, shown in Fig.~\ref{fig:Higgs_width}, increases very rapidly in this region and become comparable to its mass above 500~GeV.

\section{SIGNAL AND BACKGROUND FOR THE HIGGS DECAYING TO FOUR LEPTONS}

As the Higgs production cross-section is very small compared to the expected QCD background, rare and clean signatures based on final states with weak bosons, photons and leptons constitute the main channels for discovery. In this context, the process $\text{H} \rightarrow$ \ZZ, with the Z's decaying into leptonic final states (electrons and muons) provides a clean signal with a narrow peak on top of a relatively smooth background. If the Higgs mass is above 180~GeV, this is the so called ``golden-channel'', with two on-shell Z-bosons.

Since all the decay products are reconstructed, and considering the excellent lepton energy resolution of the ATLAS detector, this channel can give a precise measurement of the Higgs mass. For light Higgs scenarios - below 200~GeV, however, the width of the peak is dominated by the experimental resolution and no information can be extracted about the natural width of the particle.

The main background for this process is the non resonant \ZZ $\rightarrow 4$l, which has basically the same characteristics as the signal. In addition, any QCD process producing 4-lepton final states with significant transverse momenta ($p_T$) must be considered. Examples for such process are \tt\ and Z~+~jets, in particular \Zbb. Processes with 3 leptons in the final state plus one ``fake lepton'', like WZ $\rightarrow 3$l + X were found to give a negligible contribution to the search for the signal.

The cross-sections of the processes mentioned, requiring 4-leptons within the detector acceptance (transverse momentum above 5~GeV and pseudo-rapidity $\eta$ within $\pm$ 2.5), are summarized in Table~\ref{tab:xsec}.

\begin{table}[!h]
\begin{center}
\caption{Cross-sections at Next-to-Leading-Order for the signal and the main backgrounds, requiring 4-leptons with $p_T > 5$~GeV and $|\eta| < 2.5$.}
\begin{tabular}{lr}
\hline 
\textbf{Process} & \textbf{Cross-section} \\
\hline 
H $\rightarrow$ 4l (120 - 200 GeV) & 1.6 - 15.5 fb \\
\ZZ $\rightarrow$ 4l & 57.2 fb \\
\Zbb $\rightarrow$ 4l + X & 812 fb \\
\tt $\rightarrow$ 4l + X & 6064 fb \\
\hline
\end{tabular}
\label{tab:xsec}
\end{center}
\end{table}

\section{THE ATLAS DETECTOR \label{sec:ATLAS_detector}}

\begin{figure}[!h]
\centering
\includegraphics[width=0.7\textwidth]{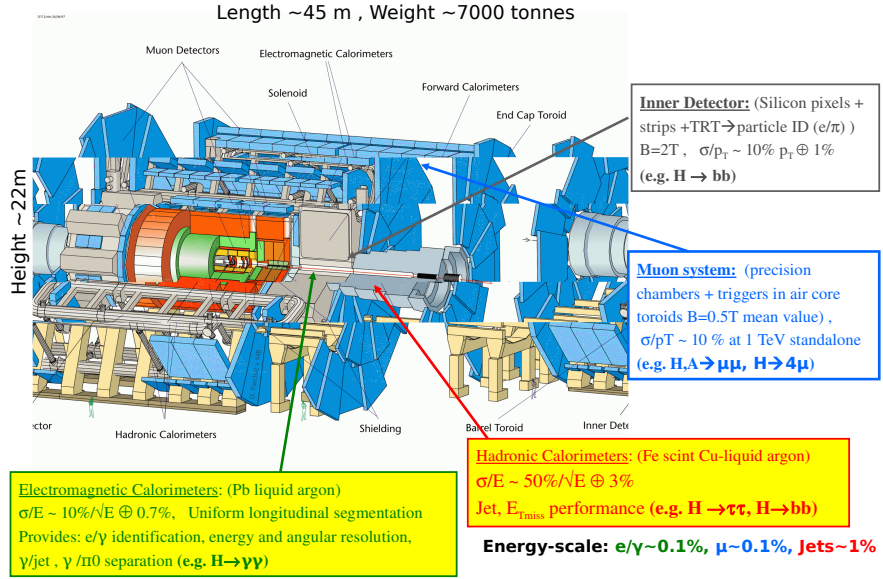}
\caption{Ilustration of the ATLAS detector.} 
\label{fig:ATLAS_detector}
\end{figure}

ATLAS is a general purpose experiment installed at the LHC collider, designed to measure and reconstruct muons, electrons, photons, taus and jets. An ilustration of the detector and some features are presented in Fig.~\ref{fig:ATLAS_detector}. Detailed information can be found at \cite{ATLAS_TDR}.

Excellent lepton reconstruction efficiency and energy resolution are essential for the H $\rightarrow$ 4l channel. An overall efficiency of the order of 95\% is foreseen for muons, combining the Inner Dectector and the Muon Spectrometer information, as described in \cite{STACO_NIM}. The combination ensures very low fake rates, below the per-mille level, and can also increase the momentum resolution with respect to the Inner Detector measurement, mainly for $p_T$ above 50 GeV.

\begin{figure}
  \label{fig:reconstruction_efficiencies}
   \subfigure[]{\includegraphics[width=0.45\textwidth]{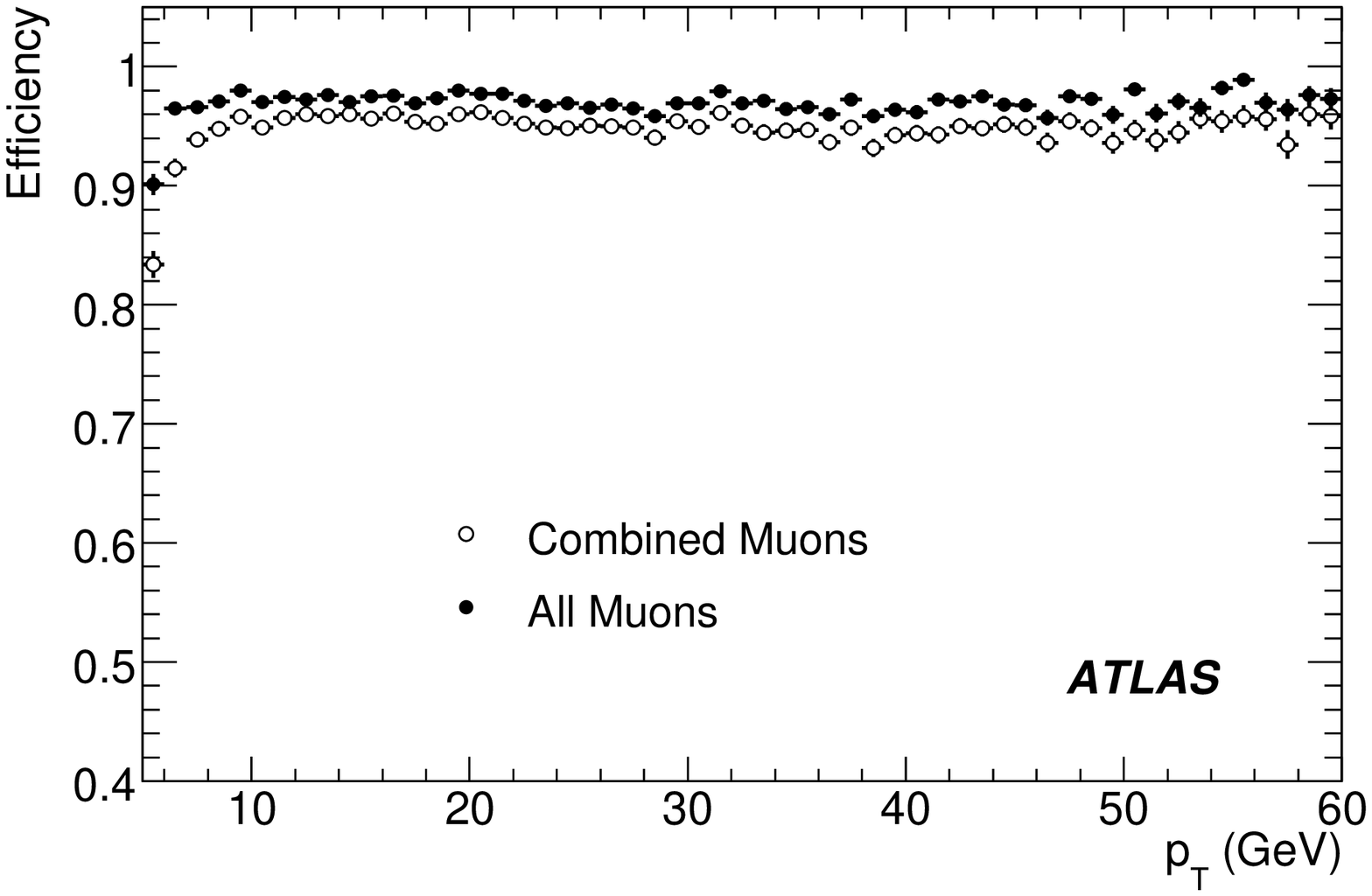}}
   \subfigure[] {\includegraphics[width=0.45\textwidth]{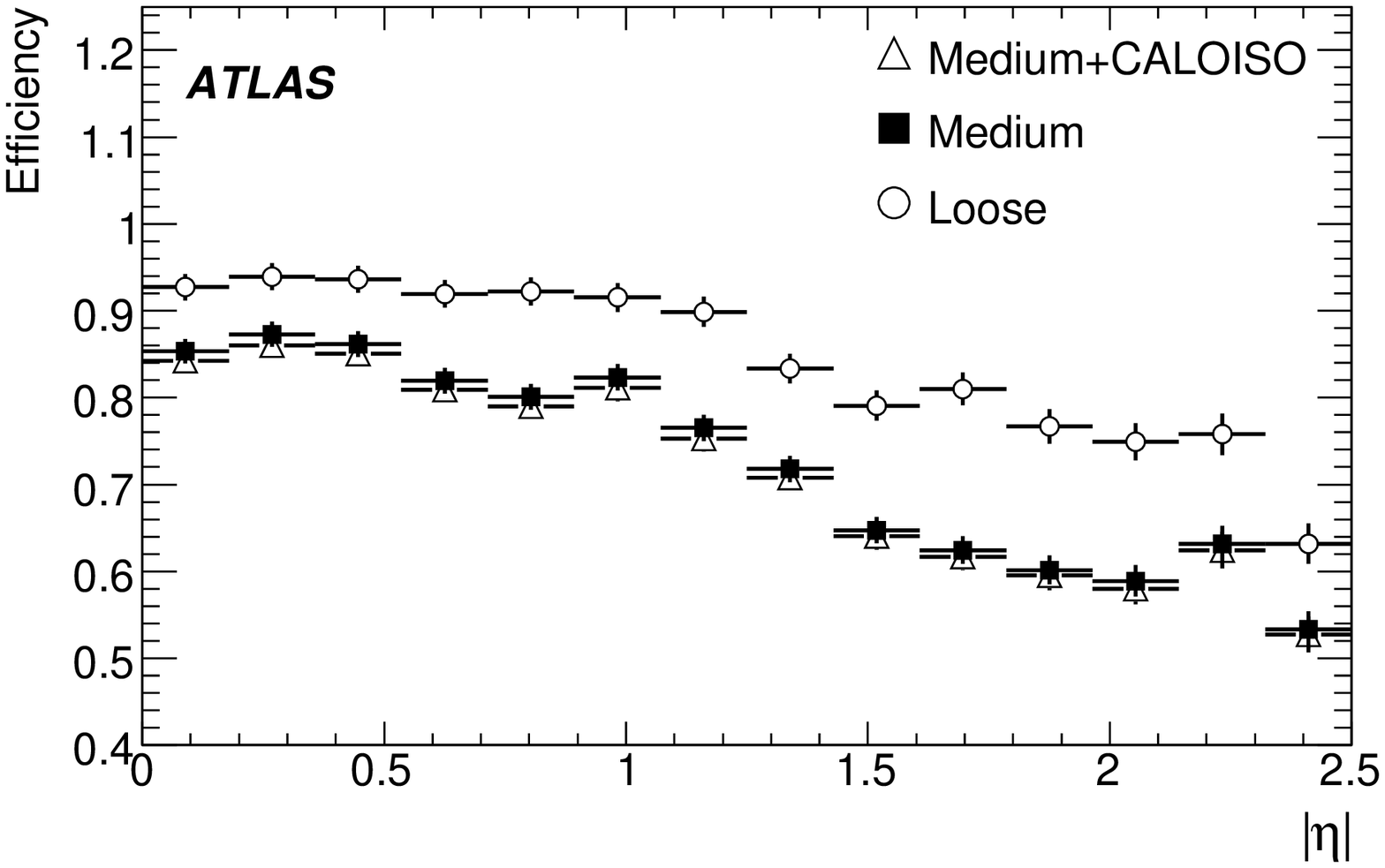}}
 \caption{(a) Muon identification efficiencies vs. $p_T$. (b) Electron identification efficiencies vs. $\eta$. Combined muons and electrons defined as \emph{Medium + CALOISO} were used in the analysis.}
\end{figure}

Electron reconstruction and identification is done by matching Inner Detector tracks with clusters from the electromagnetic calorimeter. Rejection against QCD jets is obtained by selecting clusters according to shower shape and isolation criteria. Figure~\ref{fig:reconstruction_efficiencies} shows the electron and muon reconstruction efficiencies for the definitions used in the analysis as a function of transverse momentum and pseudo-rapidity.

\section{ANALYSIS STRATEGY AND RESULTS}

The analysis consists in finding evidence for the signal by triggering and selecting apropriate events, reducing the QCD background and reconstructing the Higgs mass. Triggers for high-$p_T$ leptons (around 20 GeV) and di-leptons (above 10 GeV each) are used, with an overall efficiency around 80\% for single leptons above the threshold and greater than 95\% for the H $\rightarrow 4$l events.

\begin{figure}[!h]
 \subfigure[] {\includegraphics[width=0.45\textwidth]{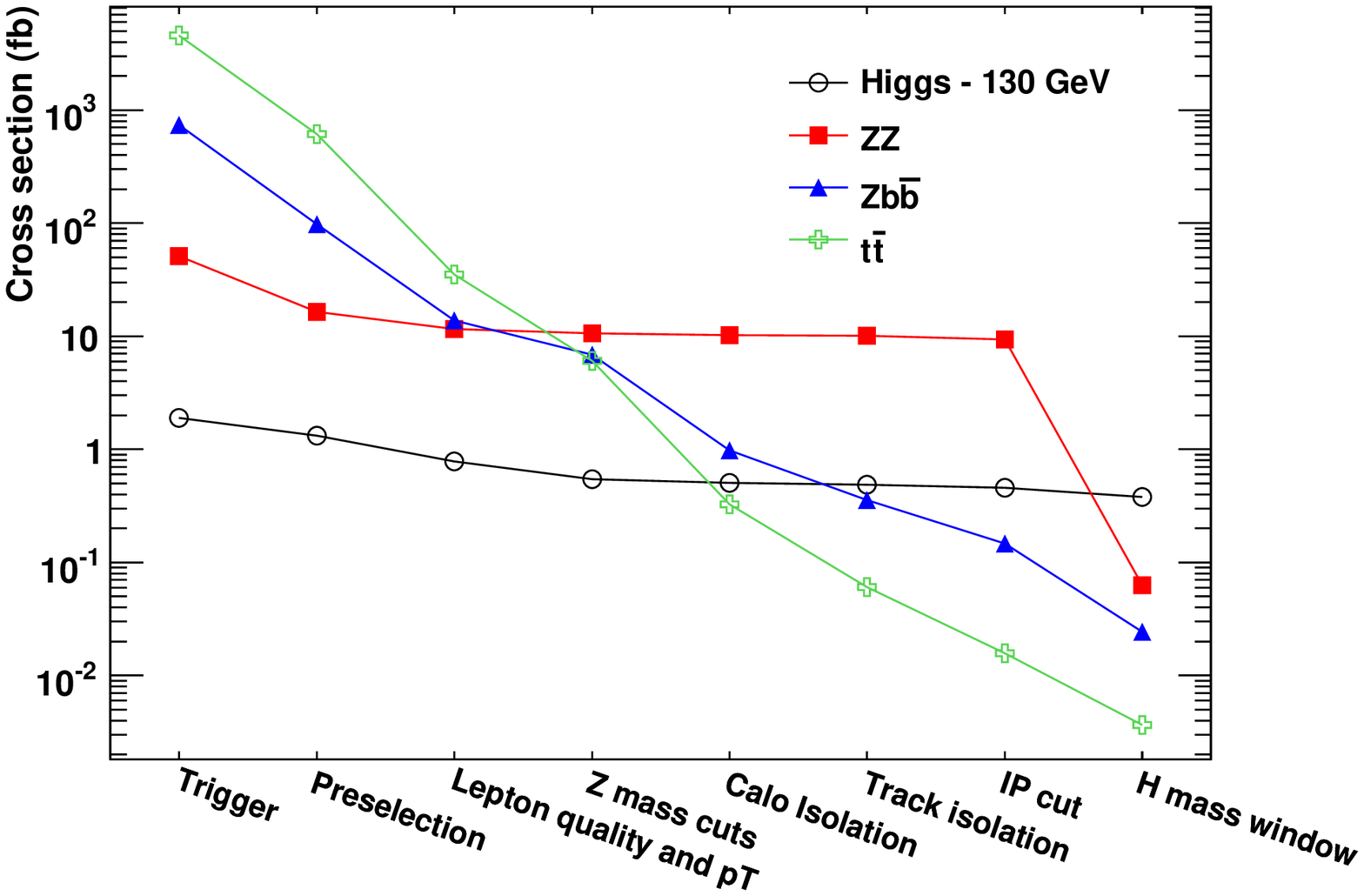}
\label{fig:xsec_cutflow}}
 \subfigure[]{\includegraphics[width=0.45\textwidth]{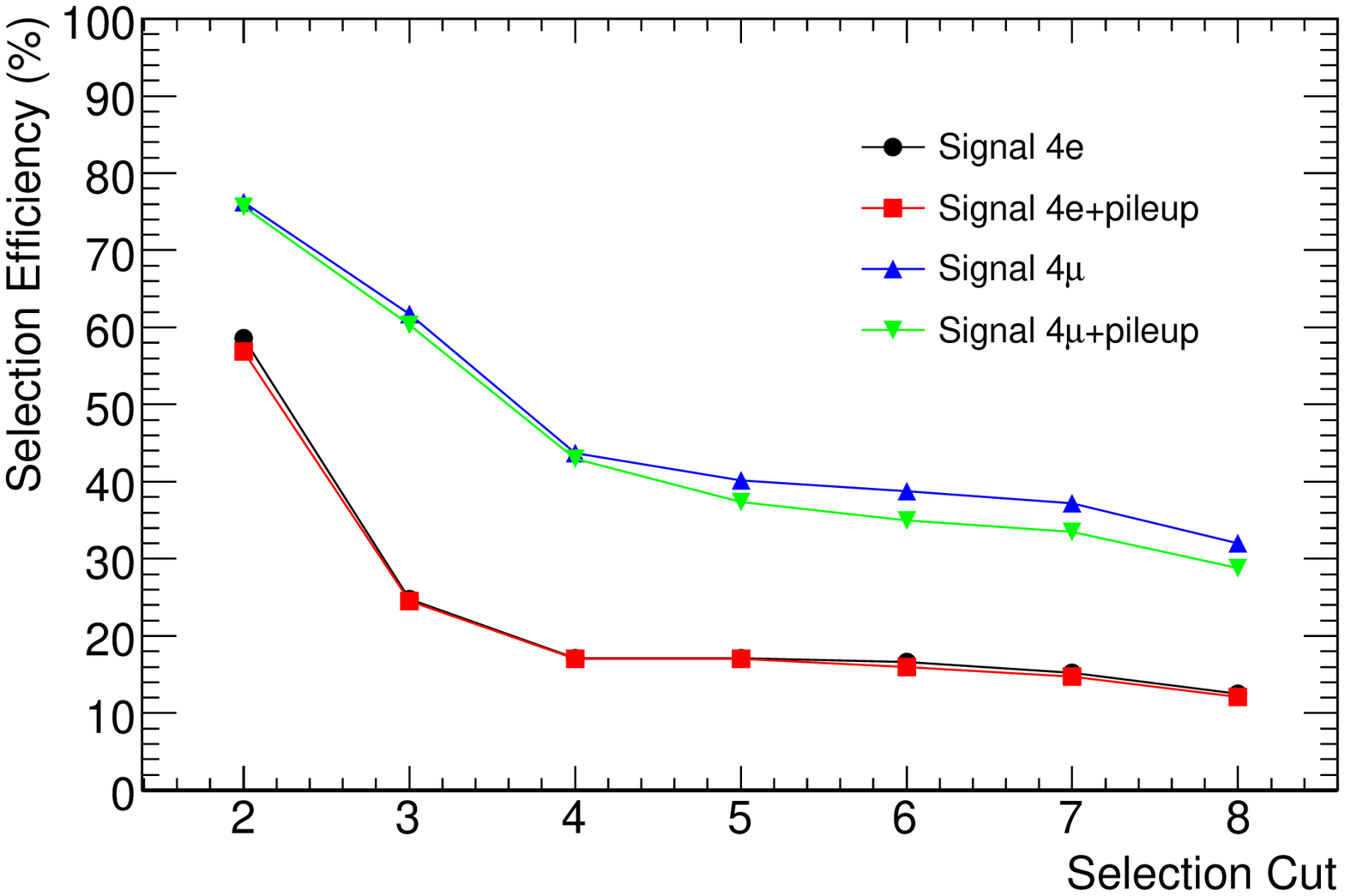}
\label{fig:selection_eff_pileup}}
\caption{(a) Cross sections after applying each analysis cut for signal and background in the 2e2$\mu$ channel. (b) Signal selection efficiency ($m_H = 130$ GeV) after applying each analysis cut for 4e and 4$\mu$ channels, including pile-up effects (see text).}
\end{figure}

A cut-based analysis was performed, to bring the level of the reducible backgrounds (in particular \Zbb) below one third of the ZZ level, as a protection against uncertainties on background yields. The events selected for the analysis were required to have at least 4 reconstructed leptons with $p_T > 5$~GeV and $| \eta | < 2.5$, as imposed in the event generation. Additional requirements on the lepton reconstruction quality and tranverse momenta are applied, exploiting the fact that leptons from bottom decays have lower $p_T$. The background is further reduced with cuts on the di-lepton invariant mass - a window around the Z mass for one lepton-pair and a lower bound for the other, calorimeter and track isolation against QCD jets, and impact parameter (IP) - to remove leptons coming from a displaced vertex. Figure~\ref{fig:xsec_cutflow} shows the cross-sections for signal and the main backgrounds after the application of each analysis cut.

Considering an instantaneous luminosity of $10^{33}\,\text{cm}^{-2}\,\text{s}^{-1}$, the effect of pile-up was investigated, including 2.3 events per bunch-crossing on LHC and thermalized slow neutrons and low energy photons escaping for the calorimeters. Figure~\ref{fig:selection_eff_pileup} shows the selection efficiency for the signal in the 4e and 4$\mu$ channel with and without pile-up effects.

\begin{figure}[!h]
 \subfigure[]{\includegraphics[width=0.4\textwidth]{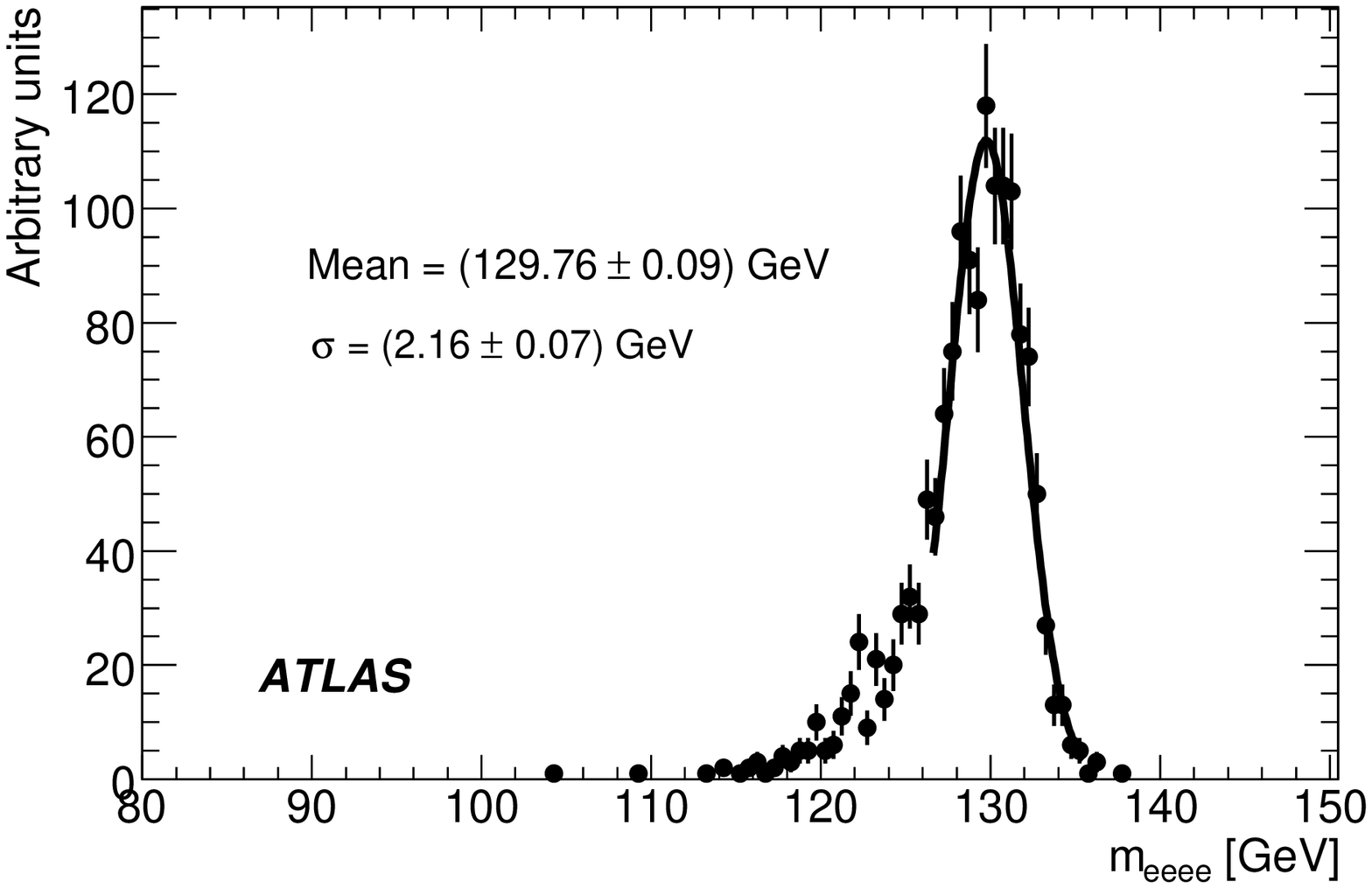}
 \label{fig:m4e}}
 \subfigure[]{\includegraphics[width=0.4\textwidth]{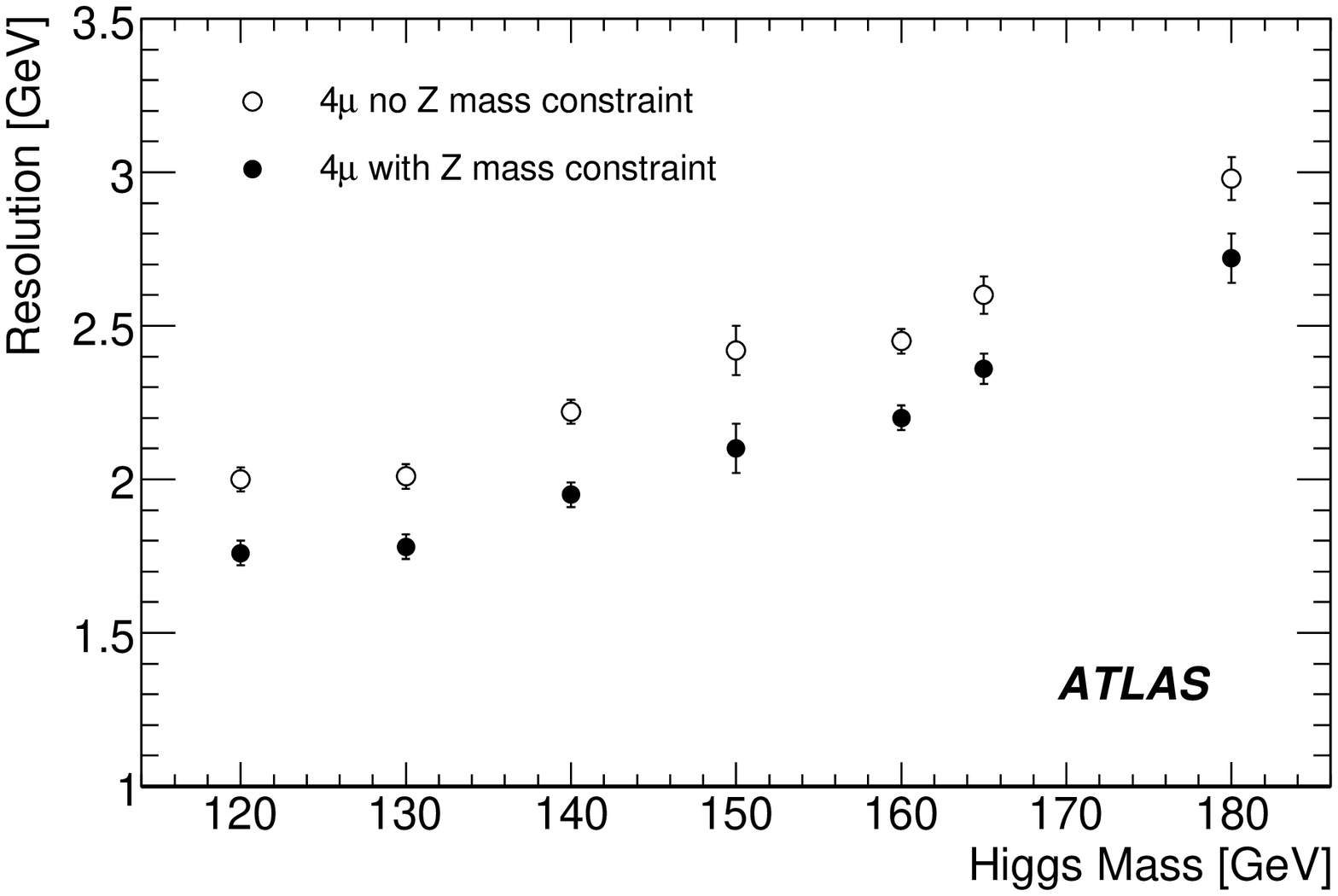}
 \label{fig:mresolution_4mu}}
 \caption{(a) Reconstructed H(130~GeV) $\rightarrow$ 4e mass after application of a Z mass constraint fit. (b) H $\rightarrow$ 4$\mu$ mass resolution with and without the application of a Z mass constraint fit.}
\end{figure}

\begin{figure}[!h]
\includegraphics[width=0.55\textwidth]{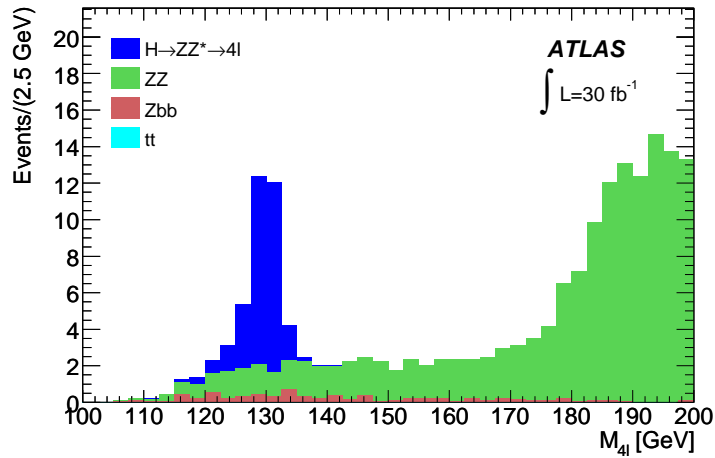}
\vspace{-0.7cm}
\caption{Reconstructed 4-lepton invariant mass applying the Z mass constraint fit, in case of a 130 GeV Higgs for a luminosity of 30 fb$^{-1}$. \label{fig:m4l}}
\end{figure}

The resolution on the measurement of the Higgs mass can be further improved by applying a constraint fit to the di-lepton pair which is considered to form the on-shell Z resonance. Fig.~\ref{fig:m4e} show the mass peak in the 4-electron channel after applying the fit, and Fig.~\ref{fig:mresolution_4mu} shows the gain provided by the fit in the 4-muon channel as a function of the Higgs mass. For a luminosity of 30~fb$^{-1}$, and combining the channels (4e, 4$\mu$ and 2e2$\mu$), one expects a clear peak in the 4-lepton invariant mass in case of a 130~GeV Higgs scenario, as illustrated in Fig.~\ref{fig:m4l}.

\newpage

\section{CONCLUSIONS}

The analysis strategy and the selection efficiencies for the Higgs decays into four lepton final states (electrons and muons) have been presented, together with the expected resolution on the mass. This proceeding is based on the ATLAS Computing Software Commissioning exercise. The notes, including the expected significance estimation, will soon be published.

%

\end{document}